\documentclass[a4paper,fleqn]{cas-sc}

\usepackage[authoryear, round, sort&compress]{natbib}
\usepackage{graphicx}%
\usepackage{multirow}%
\usepackage{amsmath,amssymb,amsfonts}%
\usepackage{amsthm}%
\usepackage{mathrsfs}%
\usepackage[title]{appendix}%
\usepackage{xcolor}%
\usepackage{textcomp}%
\usepackage{manyfoot}%
\usepackage{booktabs}%
\usepackage{algorithm}%
\usepackage{algorithmicx}%
\usepackage{algpseudocode}%
\usepackage{listings}%
\usepackage{tabularx}%
\usepackage{lscape}%
\usepackage{float}
\usepackage{pdflscape}
\usepackage{longtable}
\usepackage{multirow}
\usepackage[utf8]{inputenc}
\usepackage{url}
\usepackage{float} 
\usepackage{placeins}
\usepackage{caption}

\def\tsc#1{\csdef{#1}{\textsc{\lowercase{#1}}\xspace}}
\tsc{WGM}
\tsc{QE}
\tsc{EP}
\tsc{PMS}
\tsc{BEC}
\tsc{DE}

\begin{document}
\let\WriteBookmarks\relax
\def\floatpagepagefraction{1}
\def\textpagefraction{.001}
\shorttitle{Delays and Deferrals in Nuclear Waste Disposal}

\shortauthors{Awawda \& Wimmers}

\title [mode = title]{Delays and Deferrals in Nuclear Waste Disposal: A Stochastic Analysis of Funding Shortfalls of Germany's Waste Fund KENFO}                 

\author[1]{Mahdi Awawda}[orcid=0009-0006-1889-7911]
\author[1,2]{Alexander Wimmers}[orcid=0000-0003-3686-0793]
\cormark[1]
\ead{awi@wip.tu-berlin.de}

\address[1]{Workgroup for Infrastructure Policy (WIP), Technische Universität Berlin, Straße des 17. Juni 135, 10623 Berlin, Germany}
\address[2]{Energy, Transportation, Environment Department, German Institute for Economic Research (DIW Berlin), Anton-Wilhelm-Ano-Straße 58, 10117 Berlin, Germany}

\begin{abstract}
Germany is  tasked ensuring the safe and final storage of high-level radioactive waste in a deep geological repository. Since 2022, the ambitious target year of 2031 to identify a suitable location for such a site has been deferred by most public actors. The target year was pushed back by several decades to 2046 or even 2068, consequently delaying the completion of all waste management activities well into the 22nd century. Most radioactive waste management activities in Germany are funded via the external fund KENFO that was initiated with an initial endowment of €24.1 bn. in 2017. KENFO hopes to achieve average returns on invest (ROI) of 3.7\% over the coming decades to ensure that sufficient funds remain. However, the delays in the current process will likely result in overall cost increases. Thus, in this analysis, we conduct a stochastic analysis of the potential delays in the site selection procedure and their corresponding cost effects to assess whether KENFO's target ROI will suffice for the long-term funding requirements. We find that even under optimistic assumptions, KENFO's ROI would have to be increased to at least 5.91\%, up to 6.63\%. Alternatively, lump sum injections of up to €31.07 bn. as of 2024 could reduce funding shortfall risks. We conclude that in order to minimize the financial burden on future generations, German policymakers must address this issue of potential funding shortfalls proactively, either by reducing costs, via, e.g., delay minimization, or by increasing revenues, via, e.g., capital injections.
\end{abstract}


\begin{highlights}
\item Stochastic assessment of delays in German nuclear waste disposal
\item Analysis of financial implications of waste fund KENFO
\item Determination of minimum ROI or cash injections to ensure sufficient longterm funding
\end{highlights}

\begin{keywords}
nuclear waste management \sep German waste fund \sep stochastic analysis \sep KENFO \sep waste disposal process \sep funding shortfalls \sep Monte Carlo simulation
\end{keywords}

\maketitle
\section{Introduction} \label{sec:introduction}

On 15 April 2023, Germany closed its last three nuclear power plants. Since then, the focus of the German nuclear industry and regulators has shifted towards the  back-end of the nuclear supply chain, i.e., the decommissioning of closed power plants, and the disposal of radioactive wastes \citep{hirschhausen_ruckbau_2023}.

In 2013, after the 2011 decision to close all nuclear power plants by the end of 2022, a commission was established to evaluate a reorganization of the German nuclear waste management. In 2016, it concluded that a new law was to be established by which the German state was to identify a location for a high-level radioactive waste (HLW) repository with "the best possible safety" \citep{endlagerkommission_abschlussbericht_2016}. With this recommendation came the 2017 amendment of the German Site Selection Act (Standortauswahlgesetz, StandAG) and the creation of the current governance structure (Section \ref{sec:governance}). The current plan envisions the selection of a suitable location for an HLW repository by 2031. Following this, a repository would be constructed, wastes would be stored therein, and it would be sealed, with HLW disposal concluding before the turn of the century. The repository shall ensure the safe storage of wastes for at least one million years \citep{smeddinck_inter-_2016, smeddinck_lernende_2022, esk_verlangerte_2023}. Notably, the 2016 report found that the target year of 2031 was not achievable but concluded that all actors should aspire to minimize delays \citep{endlagerkommission_abschlussbericht_2016}. Regardless, this target was publicly communicated -- and has led to expectations that this target was in fact achievable \citep{rohlig_zum_2023, wimmers_finanzierungsfragen_2025}.

In parallel, a second commission was tasked with developing policy recommendations to ensure that all activities of the back-end were sufficiently funded without risking the bankruptcy of the operators of the power plants. The commission suggested transferring the responsibility for nuclear waste management to the state by creating an external fund with payments from the commercial operators \citep{kfk_verantwortung_2016}. So, in 2017, the German waste management fund KENFO ("Fonds zur Finanzierung der kerntechnischen Entsorgung") was created with an initial endowment of €\textsubscript{2017}23.6 bn. paid by the operators with which all future liabilities and responsibilities were transferred to the German state. KENFO is tasked with funding most nuclear waste management activities in Germany, while decommissioning remains the operators' task \citep{hirschhausen_ruckbau_2023}, see Figure \ref{fig:organization} and Table \ref{tab:KENFO_expenses}. The fund shall generate sufficient returns on its investments (ROI) to provide annual funding for ongoing activities, and to ensure that future taxpayers are not burdened with financial liabilities in addition to the substantial physical responsibilities of actually storing the waste \citep{kenfo_geschaftsbericht_2025}.

KENFO's initial fund volume was based on a 2015 assessment by consulting firm Warth \& Klein Grant Thornton\footnote{Since 2022, the company is called Grant Thornton, see \url{https://www.grantthornton.de/en/about-us/history/}, last accessed on 08-10-2025.} (WKGT) that was based on the provisions that had been made by the operators for a repository based on the now excluded Gorleben site, see Section \ref{subsec:process}. In this report, the authors assumed that the HLW repository would be sealed by the year 2098 \citep{wkgt_gutachtliche_2015}.

However, the envisioned target year of 2031 for the site identification has been publicly delayed by the responsible actors. In 2022, BGE (Bundesgesellschaft für Endlagerung), the state-owned company tasked with identifying a suitable site, published a report according to which a site selection was possible in 2046, at the earliest \citep{bge_zeitliche_2022}, and in 2024, a report issued by the regulator BASE (Bundesamt für die Sicherheit der nuklearen Entsorgung) announced that under the current design, the process would not be completed before 2074 \citep{krohn_unterstutzung_2024}.

If these delays actually occured, they would push the closure of a final repository well into the 22nd century \citep{ott_fur_2024}. In addition to regulatory, safety-related, and organizational challenges, this has implications for the provision of sufficient funds from KENFO \citep{irrek_kosten_2023}. So far, despite the delay of the German site selection procedure becoming an accepted reality, the implications for the provision of sufficient funding via KENFO remain neglected -- amidst a general lack of interest of the German public for these challenges \citep{ott_fur_2024, wimmers_finanzierungsfragen_2025}.

Thus, in this work, we provide a stochastic assessment of KENFO's required ROI targets or state-funded capital injections to avoid future funding shortfalls based on a stochastic Monte Carlo simulation. We show that, based on the available cost data, KENFO's own ROI targets, and current timelines, KENFO's funds will diminish within the next decades, and the burden of funding will likely fall on future taxpayers.

The remainder of this work is structured as follows. In Section \ref{sec:governance}, we introduce the current organization of German nuclear waste management. In Section \ref{sec:method}, we provide an overview of the cost model and the applied stochastic simulation as well as the selected scenario bundles. Section \ref{sec:results} provides the results of the simulation, while Section \ref{sec:discussion} discusses the implications and limitations of this analysis. Section \ref{sec:conclusion} concludes.






\section{Organization Nuclear Waste Management in Germany} \label{sec:governance}

\subsection{Actors} \label{subsec:actors}

\begin{figure}[h]
    \includegraphics[width=8cm]{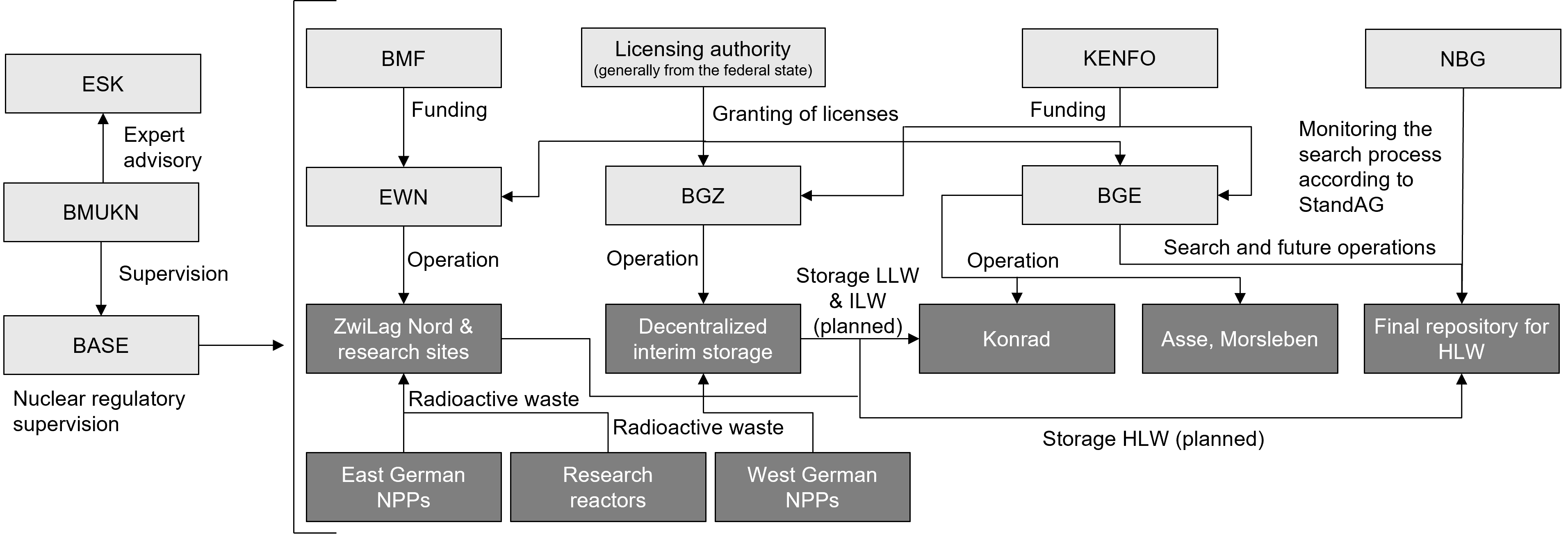}
    \centering
    \caption{The organization of nuclear waste management in Germany. Adapted from \citet{hirschhausen_ruckbau_2023}.}
    \label{fig:organization}
\end{figure}

The reorganization of nuclear waste management in Germany created several new actors. The current organizational structure is shown in Figure \ref{fig:organization}. In the following, only the most relevant actors will be briefly described following \citet{wimmers_decommissioning_2023,wimmers_entsorgung_2025}.

BASE is tasked with the regulatory oversight of all nuclear facilities in Germany, including waste management facilities, and oversees the process of site selection. It is in turn regulated by the Federal Ministry for the Environment, Climate Protection, and Nuclear Safety (BMUKN).

Radioactive waste generated at the nuclear power plants in the former GDR, Rheinsberg and Greifswald, as well as wastes from research facilities, is stored at the "Zwischenlager Nord" and at the research facilities themselves, i.e., Karlsruhe and Jülich. These storage sites are managed by EWN (Entsorgungswerke NORD), a state-owned company tasked with the interim storage of these wastes and the decommissioning of the former GDR plants and closed research reactors. EWN is funded through the state budget via the Federal Ministry of Finance (BMF).

Radioactive waste stemming from the West German power plants is stored at 17 interim storage facilities, of which two---Ahaus and Gorleben---are so-called centralized facilities. The others are located at the sites of the former power reactors. The facilities house low and medium-level wastes (LILW) as well as HLW. These facilities are owned and operated by the Federal Company for Interim Waste Storage (BGZ, Bundesgesellschaft für Zwischenlagerung) whose activities are funded via KENFO, see Table \ref{tab:KENFO_expenses} in Section \ref{subsec:KENFO}.

The Federal Company for Final Waste Storage (BGE, Bundesgesellschaft für Endlagerung) is tasked with constructing and operating the final repository for LILW, Konrad, due to begin operations "in the early 2030s" \citep{bge_endlager_2024}[p.6] after substantial delays, the closure of the former GDR repository Morsleben, and the clean-up at Asse II. Furthermore, BGE is responsible for the selection of a suitable location for an HLW repository, and will be tasked with its construction, operation, and maintenance. The site selection process is monitored by the NBG (Nationales Begleitgremium), while the Commission on (Nuclear) Waste Disposal (ESK, Entsorgungskommission) advises the BMUKN.

\subsection{Process of Final Site Repository Selection} \label{subsec:process}

Since the mid-1970s, plans had been under development to house interim and final storage facilities for different waste types, a reprocessing facility, and a fuel fabrication plant near the small village of Gorleben in Lower Saxony (at the time, this was close to the "German-German border" of the FRG and GDR). However, the site became one of the symbols of the growing Anti-Nuclear protests in West Germany, and these plans were swiftly abandonded \citep{tiggemann_elephant_2019}. From 1979 onward, however, the site was considered as a potential location for an HLW repository, and explorations began the same year \citep{bge__2020}.

In the year 2000, the exploration of Gorleben was halted for a period of three to ten years (Gorleben moratorium) and a commission (Arbeitskreis Auswahlverfahren Endlagerstandorte, AkEnd) conducted an investigation \citep{thomauske_zeitbedarf_2016}. The final report was published in 2002 with many recommendations that are also part of the current legislation, such as the "step-by-step process" and the inclusion of the affected local population, but the propositions were not enacted on due to the utilities' refusal to fund another selection procedure and the German states' refusal to host repositories \citep{tiggemann_elephant_2019}, and more generally, a lack of "political will" \citep{blowers_legacy_2017}[215, as cited by \citet{tiggemann_elephant_2019}[p.80]]. 

In 2011, the state of Baden-Württemberg initiated the restart of a site selection process which led to the introduction of the legislation mentioned in Section \ref{sec:introduction}. The amendment of the StandAG in 2017---after the commission had published its final report \citep{endlagerkommission_abschlussbericht_2016}---introduced separate paragraphs to deal with the Gorleben site that would ensure the location was treated like any other. In 2020, BGE announced that after completing the first phase of the site selection procedure (see below), Gorleben is no longer considered as a potential location \citep{bge__2020}.

The current site selection procedure, legislatively defined in the StandAG, defines three major steps for the site selection, see Table \ref{tab:phases}. The first phase ("Identification of Potential Regions") is divided into two steps and was envisioned to be completed in 2022. The first step is the identification of "partial regions" ("Teilgebiete"), completed in September 2020, and the second is the identification of potential regions for further analysis (ongoing). The second phase is the overground exploration of the potential regions, consisting of four steps, and originally planned to conclude with several suitable locations for underground exploration in 2027. Finally, the third phase was envisioned to end with a recommendation for a suitable location in 2031 to be voted on by the Bundestag, the German parliament \citep{krohn_unterstutzung_2024}. The plan was deemed to be unrealistic by the commission as early as 2016 \citep{endlagerkommission_abschlussbericht_2016}.

Selecting a suitable location is only the first step of a long-term infrastructure project. After establishing the legal certainty of the site selection procedure, the repository will have to be licensed and constructed -- these steps could require an additional 30 to 60 years until they are completed \citep{thomauske_zeitbedarf_2016}. Refer to Figure \ref{fig:standAG_SOLL} in the appendix for an overview of the disposal plan for HLW as envisioned in the StandAG.

In 2022, BGE stated that a site selection was to be expected for 2046, the earliest, or even later, up to 2068 \citep{bge_zeitliche_2022}. Table \ref{tab:phases} shows BGE's 2022 estimates for the respective steps. A report commissioned by BASE places the most optimistic estimate to 2074 \citep{krohn_unterstutzung_2024}. These delays push the construction and operation of the repository well into the 22nd century and result in several challenges.

\begin{table}[h!]
\centering
\scriptsize
\begin{tabular}{p{3cm}p{3cm}p{3cm}p{3cm}} \hline
\multicolumn{2}{c}{\textbf{Phase}} & \textbf{Description}                              & \textbf{Schedule (planned as of 2022)}  \\ \hline
I           & Step 1               & Identification of "Teilgebiete"                   & Completed in September 2020  \\
            & Step 2               & Identification of regions for further exploration & Scheduled for completion in second half of 2027 \\
II           & Scenario A           & Overground exploration of 6 regions               & 10 years  \\
            & Scenario B           & Overground exploration of 12 regions              & 12 years  \\
III         & Scenario A           & Underground exploration via boreholes             & 5 to 6 years  \\
            & Scenario B           & Underground exploration via exploratory mines     & 12 to 23 years \\ \hline  
\end{tabular}
\caption{The three phases of the site selection procedure and estimated durations according to \citet{bge_zeitliche_2022}.}
\label{tab:phases}
\end{table}

First, the delay in the site selection results in extended interim storage periods for HLW. Currently, radioactive waste is stored at 15 decentralized and two centralized interim storage facilities operated by BGZ and EWN. The storage containers are licensed for 40 years and might have to be refurbished or replaced, which in turn necessitates suitable infrastructure for HLW handling, such as sufficiently dimensioned hot cells, which does not exist in Germany as of today. \citep{wimmers_entsorgung_2025}

Second, it must be ensured that institutional knowledge remains available. With Germany having ended the operations of its commercial nuclear power plants, know-how is beginning to decline, and many specialists are on the verge of retirement, and might have even passed away until the repository becomes available. Ensuring the availability of skilled personnel is a major challenge. \citep{barenbold_decommissioning_2024}

Finally, societal challenges relate first to the current lack of interest regarding the topic, and potential conflicts that will surely arise once locations are chosen for further exploration in the coming years -- legal disputes will delay the process even more. Furthermore, the delays mean a transfer to liabilities onto future generations to deal with. This goes not only for the physical implementation of a repository, but also for the provision of sufficient funds for the decades to come. \citep{wimmers_finanzierungsfragen_2025}

The remainder of this work is concerned with the latter challenge of sufficient funding.







\subsection{Introducing KENFO and the Organization of German Waste Management Funding} \label{subsec:KENFO}

KENFO was created in 2017 with an initial endowment of €23.6 bn. Following the "Law for the Reorganization of the Financing of Nuclear Waste Management" (Gesetz zur Neuordnung in der kerntechnischen Entsorgung, VkENOG), this sum was paid by the utilities operating or owning the West German nuclear power plants, i.e., E.On (the owner of nuclear operator PreussenElektra), RWE, Vattenfall, and EnBW. The payment consisted of €17.4 bn based on the utilities' provisions for waste management obligations, and an additional risk surcharge of €6.2 bn, the payment of which rendered any future financial liabilities of the utilities null and void. With this payment, the responsibility for managing nuclear wastes in Germany became the state's responsibility. Decommissioning remains the responsibility of the utilities. Research reactors and former GDR plants are decommissioned by German state actors \citep{barenbold_decommissioning_2024}, see Figure \ref{fig:organization}.

\subsubsection{KENFO's Funding Responsibilities}

Apart from payments to the respective local community support funds for Asse, Morsleben, and Salzgitter (Gorleben), the clean-up of the Asse II mine as well as the closure of the former GDR repository for LILW in Morsleben, all waste management activities are funded by KENFO. Activities are paid for by BMUKN which is subsequently reimbursed by KENFO following dedicated laws and ordinances. The actual and planned expenses of KENFO from 2019 to 2026 are shown in Table \ref{tab:KENFO_expenses}. Note that for 2024 to 2026, data has not been updated, and since 2023, BMUKN estimates costs only for the next year instead of up to four years into the future.

\begin{table}[]
\scriptsize
\begin{tabular}{p{2cm}p{2cm}p{0.5cm}p{0.5cm}p{0.5cm}p{0.5cm}p{0.5cm}p{0.5cm}p{0.5cm}p{0.5cm}} \hline
\multirow{2}{*}{\textbf{Task}}    & \multirow{2}{*}{\textbf{Reimbursed via}} & \multicolumn{7}{l}{\textbf{Expenses in € m. (nominal)}}                                                                                                                                                                                                               & \multicolumn{1}{l}{}                 \\
                                  &                                                    & 2019             & 2020                                 & 2021                                 & 2022                                 & 2023                                 & 2024 (planned)                       & 2025 (planned)                       & 2026 (planned)                       \\ \hline
Project   Konrad                  & EndlagerVIV                                        & 300              & 367.4                                & 295.12                               & 323.47                               & 362.42                               & 365                                  & 421.66                               & 325.82                               \\
Site   Selection Procedure        & StandAG                                            & 43.2             & 24.4                                 & 41.55                                & 36.91                                & 54.14                                & 50                                   & 67.85                                & 76.88                                \\
Project   Gorleben                & StandAG                                            & 15               & 15.34                                & 14.2                                 & 14.45                                & 21.55                                & 20                                   & 34.56                                & 25.41                                \\
Administrative   expenses of BASE & EndlagerVIV \& StandAG                             & 34.65            & 41.63                                & 45.12                                & 54.41                                & 63.49                                & 72.74                                & 74.34                                & 63.25                                \\
Interim   storage                 & EntsorgÜG                                          & 401.71           & 415.7                                & 413.87                               & 353.83                               & 430.58                               & 430                                  & 535.44                               & 541.67                               \\
Product   control activities      & Fees charged from waste   producers                & 10.4             & 14.3                                 & 22.47                                & 26.32                                & 27.47                                & 15                                   & 22.78                                & 29.46                                \\ \hline
\multicolumn{2}{l}{\textbf{Total reimbursable expenses}}                               & \textbf{804.96}  & \multicolumn{1}{l}{\textbf{878.77}}  & \multicolumn{1}{l}{\textbf{832.33}}  & \multicolumn{1}{l}{\textbf{809.39}}  & \multicolumn{1}{l}{\textbf{959.65}}  & \multicolumn{1}{l}{\textbf{952.74}}  & \multicolumn{1}{l}{\textbf{1156.63}} & \multicolumn{1}{l}{\textbf{1062.49}} \\ \hline
Fund for   Salzgitter             & not reimbursable                                   & 0.7              & 0.7                                  & 0.7                                  & 0.7                                  & 0.7                                  & 0.7                                  & 0.7                                  & 0.7                                  \\
Fund for   Morsleben              & not reimbursable                                   & 0.4              & 0.4                                  & 0.4                                  & 0.4                                  & 0.4                                  & 0.4                                  & 0.4                                  & 0.4                                  \\
Fund for   Asse                   & not reimbursable                                   & 3                & 3                                    & 3                                    & 3                                    & 3                                    & 3                                    & 3                                    & 3                                    \\
Clean-up   at of Asse II mine     & not reimbursable                                   & 160              & 195.97                               & 174.03                               & 162.52                               & 191.53                               & 190                                  & 214.4                                & 248.73                               \\
Closure   of Morsleben repository & not reimbursable                                   & 49.8             & 67.49                                & 66.65                                & 66.8                                 & 72.35                                & 70                                   & 99.57                                & 67.12                                \\ \hline
\multicolumn{2}{l}{\textbf{Total non-reimbursable expenses}}                           & \textbf{213.9}   & \multicolumn{1}{l}{\textbf{267.56}}  & \multicolumn{1}{l}{\textbf{244.78}}  & \multicolumn{1}{l}{\textbf{233.42}}  & \multicolumn{1}{l}{\textbf{267.98}}  & \multicolumn{1}{l}{\textbf{264.1}}   & \multicolumn{1}{l}{\textbf{318.07}}  & \multicolumn{1}{l}{\textbf{319.95}}  \\
\multicolumn{2}{l}{\textbf{Total expenses}}                                            & \textbf{1018.86} & \multicolumn{1}{l}{\textbf{1146.33}} & \multicolumn{1}{l}{\textbf{1077.11}} & \multicolumn{1}{l}{\textbf{1042.81}} & \multicolumn{1}{l}{\textbf{1227.63}} & \multicolumn{1}{l}{\textbf{1216.84}} & \multicolumn{1}{l}{\textbf{1474.7}}  & \multicolumn{1}{l}{\textbf{1382.44}} \\ \hline
\end{tabular}
\caption{Actual and planned expenses for German nuclear waste management activities. Note that the term "reimbursable" means that activities are paid for by KENFO. Own compilation based on \citet{bmu_bmuv-haushalt_2019, bmu_bmu-haushalt_2020,bmuv_bmuv-haushalt_2022,bmuv_bmuv-haushalt_2023,bmuv_bmuv-haushalt_2024}. Abbreviations used: BASE = Bundesamt für die Sicherheit der nuklearen Entsorgung; EndlagerVIV = Endlagervorausleistungsverordnung; EntsorgÜG = Entsorgungsübergangsgesetz; StandAG = Standortauswahlgesetz.}
\label{tab:KENFO_expenses}
\end{table}




KENFO is tasked with providing funds for the "safe disposal of current and future radioactive wastes generated by the commercial operation of nuclear power to generate electricity in Germany" \citep{kenfo_geschaftsbericht_2025}[p.6]. What exactly this entails, e.g., whether KENFO's funds shall last until the closure of the repository, when all waste has been stored therein, or whether subsequent monitoring activities, planned for at least 500 years, shall also be funded, remains uncertain \citep{brunnengraber_fondsmodelle_2024}.

\subsubsection{The Expected Costs of German Nuclear Waste Management}

In general, determining the costs of nuclear waste management activities is challenging. First, because of the lack of global experience \citep{wimmers_disposal_2026}, and second, because of the lack of (international) regulations on what cost components to include, such as costs for research and administrative costs, public consultations, site explorations, transport of materials, the construction of the repository, future monitoring, and more. Refer to \citet{iaea_costing_2020} for an overview of potential cost components in nuclear waste management. Third, very long timeframes are subject to inherent uncertainty.

The last comprehensive German nuclear waste management cost study was published in 2015 by auditing firm WKGT. The analysis included both waste management and decommissioning costs and was based on the utilities' provisions which were in turn based on cost assumptions for a repository concept modeled on the plans for Gorleben -- a design that had by then been outdated for at least a decade. The analysis assumes the repository will close in 2098, and assumes discounted costs of approximately €\textsubscript{2014}51 bn. for the repository alone, and an additional €\textsubscript{2014}89 bn. for transportation casks, interim storage, and the Konrad repository \citep{wkgt_gutachtliche_2015}.

With the utilities' payment of the risk premium, all future liabilities have been transferred to the German state. Consequently, KENFO cannot rely on additional payments by either electricity consumers or the utilities. Furthermore, the potential future costs to be funded by KENFO are not subject to periodical scrutiny, like they are in Switzerland, where waste levies are periodically updated based on recurring cost studies \citep{brunnengraber_fondsmodelle_2024, barenbold_decommissioning_2024, wimmers_finanzierungsfragen_2025}.

KENFO must therefore provide sufficient funds for the coming decades solely through returns on its investments (ROI). KENFO began with a target ROI of 3.7\% \citep{mikus_wir_2020}, which was increased to 4.11\% in 2024, and set to 3.95\% for 2025 \citep{kenfo_geschaftsbericht_2025}. Actual ROIs were 11.1\% and 9.4\% in 2023 and 2024, respectively \citep{kenfo_geschaftsbericht_2025}, and had been above target for several years \citep{mikus_rede_2023}. In 2022, KENFO accounted for losses of 12.2\% \citep{kenfo_geschaftsbericht_2023}. These variations demonstrate the uncertainty surrounding investment portfolios that could bear the risk of not achieving their target ROIs in the future, especially in times of heightened geopolitical tensions. Refer to \citet{eckhardt_wicked_2024} and \citet{brunnengraber_generationenkapital_2024} for an assessment of KENFO's investment strategies.

Furthermore, the to-be-funded costs remain unpredictable \citep{eckhardt_wicked_2024}. First, the very long time horizon of the German waste management project bears uncertainty in itself. The above-mentioned delays in an already decade-long process will increase this uncertainty. Second, large-scale infrastructure projects in Germany, including projects in the nuclear industry, are prone to cost escalations and delays \citep{johnstone_beyond_2020, locatelli_why_2018}. For example, the Konrad mine, which shall function as a repository for LILW, is delayed by several years and will cost at least €5 bn.\footnote{The last delay and cost update was announced in 2023, see \url{https://www.heise.de/news/Atommuell-Endlager-Konrad-wird-teurer-und-kommt-spaeter-9538609.html.}, last accessed on 01-10-2025.} Third, the nuclear sector in itself is prone to underestimating costs, especially in the back-end. This is exemplarily shown by underestimated decommissioning project costs \citep{barenbold_decommissioning_2024}, and by the underestimation of interim waste storage costs in Germany \citep{irrek_kosten_2023}.

\section{Method} \label{sec:method}

In the previous sections, we established that, despite several organizational changes and delays, there has been no update on the costs for the German nuclear waste disposal process for at least a decade, and, to our knowledge, there is no publicly available inquiry of KENFO's future funding adequacy \citep{wimmers_finanzierungsfragen_2025}. Thus, this work provides a first stochastic analysis of KENFO's funding adequacy. We analyze KENFO's potential funding shortfalls under given cost estimations and target ROIs. We then calculate KENFO's minimal target ROI under different assumptions following the projected delays of the site selection process based on the assessments by \citet{bge_zeitliche_2022}. We further provide an assessment of minimum capital injections from today's state budgets that could (theoretically) ensure KENFO's funding adequacy with lower ROIs.

In the following, we provide an overview of the cost model based on \citet{wkgt_gutachtliche_2015}, and define the minimization problem. We then describe the stochastic model, and define the scenarios.

\subsection{Cost Model and Minimization Problem} \label{sec:model}

The financial assessment of Germany’s nuclear waste management is formulated here as a cost--minimization problem, drawing on the cost projections by \citet{wkgt_gutachtliche_2015} and adapting them to recent developments. To our knowledge, \citet{wkgt_gutachtliche_2015} provide the only comprehensive publicly available cost assessment for nuclear waste disposal in Germany. It estimated a total expenditure of approximately €\textsubscript{2014}140 bn. for activities from 2015 to 2099, encompassing interim storage, the development of the final repositories, their operation and closure, as well as the decommissioning of the closed power plants.


The adapted cost model structures costs into five categories: (i) interim storage, including the continued operation of decentralized facilities; (ii) siting, development and construction of the final repository; (iii) its operation, including transport of wastes and loading; (iv) closure and post-closure monitoring, and for one scenario, (v) the costs for consolidated interim storage facilities. Each of these cost components is distributed across future years, with time horizons depending on scenarios. As the site selection and construction process has already experienced substantial delays, additional cost increases must be expected due to extended storage requirements and inflationary pressures \citep{bge_zeitliche_2022, ott_fur_2024}. Other activities funded by KENFO, such as Konrad, are not adapted. We limit the analysis to parameters directly related to the HLW repository to reduce the already high degree of uncertainty.


From the perspective of an externally segregated fund such as KENFO, the adequacy of financing depends on whether the accumulated assets---originating from the initial capital transfer in 2017---can generate sufficient investment returns to meet projected expenditures. Let $C_t$ denote the cost in year $t$, and $F_t$ the fund’s available assets. After accounting for returns on investment and disbursements, adequate funding requires that for each period $t$, \(F_t \geq C_t\). This condition can be expressed as a minimization problem that determines the lowest constant annualized rate of return, or alternatively, the minimal upfront capital injection to ensure solvency over the planning horizon.

Formally, the problem is defined as follows: Given a stochastic distribution of project milestones---site selection, repository construction, commissioning, and closure---the model minimizes the required annual average return on investment (ROI), subject to the constraint that the fund balance remains greater than zero. Equivalently, if the ROI is fixed at a certain level, the optimization determines the minimum lump sum injection that would be required today (end of year 2024) to guarantee adequacy. This yields two complementary policy indicators: the required ROI threshold and a potential taxpayer-funded backstop.

Solving this minimization problem requires the integration of temporal uncertainty into cost flows. The dates of key milestones are treated as random variables, modeled via Monte Carlo simulation. Each simulated trajectory yields a unique stream of annual costs, adjusted for inflation and nuclear-specific cost escalation. For each scenario, the optimization algorithm evaluates whether the fund can cover all expenses and iteratively adjusts the ROI (or injection level) until adequacy is achieved. The procedure thus links uncertain project timelines with financial requirements, translating technical delays directly into fiscal stress on the fund.

The model is intentionally conservative in scope. It assumes no reinvestment of disbursed funds, excludes potential gains from accelerated technological innovation, and does not incorporate KENFO’s commitments to ESG criteria. Instead, it provides a lower-bound estimate of funding adequacy under uncertainty. By benchmarking against the cost baseline of the WKGT study, all while adjusting for stochastic delays, the minimization framework offers a transparent tool to assess whether KENFO’s long-term financial design is robust or whether supplementary taxpayer contributions might have to be publicly discussed. This means that the results for minimum ROIs or cash injections must be considered as minimal. So, if funding were to be actually secured, contingencies would have to be added to our results.

\subsubsection{Cost Minimization Model} \label{sec:minimization}

We define a minimization problem to solve for minimum ROI. The model is easily adaptable to calculate the minimum required lump sum injection.\footnote{The minimization problem can be rewritten to minimize the fund balance at t=1 instead of optimizing for ROI.}

\begin{equation} \label{cha7_eq2}
Objective \space Function: Minimize \space ROI
\end{equation}

\begin{equation} \label{cha7_eq3}
s.t. \space Costs_{t} \leq LiquidAssets_{t-1}
\end{equation}

\begin{equation} \label{cha7_eq4}
ROI = \left( \frac{1}{T} \sum_{t=1}^{T} \frac{FundBalance_{t}+Costs_{t}}{FundBalance_{t-1}} \right) -1
\end{equation}

\begin{equation} \label{cha7_eq5}
FundBalance_t=FundBalance_{t-1}*(1+ROI)-Costs_t
\end{equation}

\begin{equation} \label{cha7_eq6}
FundBalance_t=LiquidAssets_t+IlliquidAssets_t
\end{equation}

\begin{equation} \label{cha7_eq7}
LiquidAssets_t=(CashPercentage_t+LiquidPercentage_t )*FundBalance_t
\end{equation}

\begin{equation} \label{cha7_eq8}
LiquidPercentage_t = \begin{cases} 
      0.7 & t < \frac{T}{4} \\
      0.6 & \frac{T}{4}\leq t < \frac{T}{2} \\
      0.5 & \frac{T}{2}\leq t < \frac{3T}{4} \\
      0.4 & t \geq \frac{3T}{4}
    \end{cases}
\end{equation}

\begin{equation} \label{cha7_eq9}
    CashPercentage_t=max\{\left(0.08-\frac{0.08}{5}*(t-1)\right);0\}
\end{equation}

\begin{equation} \label{cha7_eq10}
    Costs_t \in CostProjections= \{Costs_{t=1},Costs_{t=2},...,Costs_{t=T}\}
\end{equation}

\begin{equation} \label{cha7_eq11}
Costs_t = i\sum_{s}{F_{t,s}}+ j\sum_{s}{I_{t,s}} + k\sum_{s}{D_{t,s}}
\end{equation}

\begin{equation} \label{cha7_eq12}
    \begin{split}
        F_{t,s} = Capital_{t,s}+Operation_{t,s}+Transport_{t,s}+Disposal_{t,s}+ \\
        Regulatory_{t,s}+Misc_{t,s}
    \end{split} 
\end{equation}

\begin{equation} \label{cha7_eq13}
I_{t,s} = Capital_{t,s}+Operation_{t,s}+Transport_{t,s}+ \\
Regulatory_{t,s}+Misc_{t,s}
\end{equation}

\begin{equation} \label{cha7_eq14}
    \begin{split}
        D_{t,s} = Decontamination_{t,s}+Demolition_{t,s}+Transport_{t,s}+ \\ SiteRestoration_{t,s}+Safeguarding_{t,s}+Misc_{t,s}
    \end{split}
\end{equation}

\begin{equation} \label{cha7_eq15}
    t \in \{0,1,...,T\} \in \mathbb{N}
\end{equation}

\begin{equation} \label{cha7_eq16}
    s \in \{Site_1, ..., Site_n\}
\end{equation}

\begin{equation} \label{cha7_eq17}
    i,j,k \in \{0,1\}
\end{equation}

The average annual rate of return must cover the projected costs. $ROI$ is the return on investment expressed as a percentage (Eq. \ref{cha7_eq3} and \ref{cha7_eq4}). $FundBalance_t$ is the balance of the fund at the end of year $t$ (Eq. \ref{cha7_eq5} and \ref{cha7_eq6}). The fund must cover costs until the project's end year $T$ (Eq. \ref{cha7_eq3}). $Costs_t$ are the projected costs for nuclear waste management activities in year $t$ (Eq. \ref{cha7_eq10} and \ref{cha7_eq11}). $LiquidAssets_t$ is the portion of the fund that is liquid in year $t$, consisting of easily monetizable assets, such as government and corporate bonds, REITs, and cash. $IlliquidAssets_t$ cannot be easily monetized, such as private equity, infrastructure investments, private debt, and real estate (Eq. \ref{cha7_eq6}). $LiquidPercentage_t$ gives the share of liquid assets in the total fund that changes according to the KENFO’s targets discussed in Section \ref{subsec:KENFO} (Eq. \ref{cha7_eq7} and \ref{cha7_eq8}). KENFO is planning to reduce its cash holdings from 8\% to 0\% within five years. Thus, there will be no generated yield for the respective percentage of the fund. Thus, cash holdings must be deducted from the ROI calculation. For this, a linear decrease is assumed, starting at 8\% in $t=1$, down to 0\% by $t=5$ (Eq. \ref{cha7_eq9}).

Depending on the country and its external segregated fund’s regulation, the external segregated fund might have defrayed costs of the whole process, i.e., projects relating to waste management such as a subset of those. Hence, subdividing the cost projection variable by formulating these into binary variables is particularly important.
$F_{t,s}$ are the final disposal cost projections for a given site $s$ in year $t$. Analogously, $I_{t,s}$ and $D{t,s}$ are the interim storage and decommissioning costs, respectively. Binary variables $i$, $j$, and $k$ indicate whether the cost component is covered by the fund. For KENFO, $k = 0$.\footnote{The detailed cost structure is available on GitLab (\url{https://github.com/Mahdi-Awawda/KENFO)}.}

To identify the minimum constant ROI that ensures pathwise adequacy, we embed the simulation within a bisection routine on ROI. Starting from conservative lower/upper bounds, the midpoint ROI is tested by running the full cash-flow projection across all simulated timelines; if any path violates solvency, the lower bound is raised, otherwise the upper bound is lowered, until the bracket meets a prescribed tolerance. For the lump sum minimization, ROI is fixed ex ante and the algorithm searches for the minimal initial balance (interpretable as a lump-sum capital injection today) that yields adequacy.\footnote{Capital injections from taxpayer money have been used at other waste funds, such as \pounds10 bn. that were paid to the United Kingdom's Nuclear Liabilities Fund in 2020 and 2021 \citep{wimmers_organizational_2024}.}

\subsubsection{Stochastic Model} \label{sec:montecarlo}

To incorporate timeline uncertainty into the funding analysis, we employ a Monte Carlo framework that translates risk-managed schedule slippages into annual cost flows. The risk factors, probabilities, and potential delays are based on \citet{bge_zeitliche_2022}[pp. 53-66, 87-91, 98-100] and listed in Table \ref{tab:bge_risks} in the annex. The stochastic driver is the year in which the deep geological repository completes its disposal operations. Let $T$ denote that completion year, defined as

\begin{equation} \label{cha7_eq18}
T \;=\; T_1 \;+\; T_2 \;+\; T_3 \;+\; T_{\mathrm{op}},
\end{equation}

where $T_1$ is the (deterministic) end of Phase~I, $T_2$ and $T_3$ are the durations of Phases~II and~III respectively (see Section \ref{subsec:process}), and $T_{\mathrm{op}}$ is the fixed disposal operation period. Phases~II and~III each comprise two sub-scenarios. In Phase~II (surface-based exploration), sub-scenario~A assumes an exploratory assessment of six candidate siting regions requiring a total of ten years, whereas sub-scenario~B considers ten candidate siting regions and yields a duration of twelve years. In Phase~III (subsurface exploration), sub-scenario~A assumes the use of boreholes, whereas sub-scenario~B assumes subsurface exploration by mining, with a duration of 13--23~years depending on the host rock. Since it remains uncertain whether BGE will select sub-scenario~A or sub-scenario~B for Phases~II or~III, we assume an equal probability of occurrence, i.e., 50\% for each sub-scenario (see Table \ref{tab:phases}). Consistent with the process design, we set $T_1 = 2027.5$ and $T_{\mathrm{op}} = 49$ years \citep{bge_zeitliche_2022,esk_verlangerte_2023}.

Phase durations are modeled as the sum of a baseline range and risk‐driven increments. For $j \in \{2,3\}$:

\begin{equation} \label{cha7_eq19}
T_j \sim U(a_j,b_j) \;+\; \sum_{r=1}^{n_j} Z_{j,r},
\end{equation}

with $U(\cdot,\cdot)$ a uniform law over BGE’s minimum and maximum phase durations, and $Z_{j,r}$ the realized delay from risk $r$ in phase $j$. Each risk event is characterized by an occurrence probability $P_{j,r}$ and an impact $D_{j,r}$ (in months), both drawn from the BGE risk tables’ reported bounds:

\begin{equation} \label{cha7_eq20}
P_{j,r} \sim U\!\bigl(p^{\min}_{j,r},p^{\max}_{j,r}\bigr), 
\qquad 
D_{j,r} \sim U\!\bigl(d^{\min}_{j,r},d^{\max}_{j,r}\bigr).
\end{equation}

Realizations follow a Bernoulli–uniform compound,

\begin{equation} \label{cha7_eq21}
Z_{j,r} \;=\; 
\begin{cases}
D_{j,r}, & \text{with probability } P_{j,r},\\[2pt]
0, & \text{otherwise},
\end{cases}
\end{equation}

so that the phase‐level stochastic delay is $\Delta T_j=\sum_{r=1}^{n_j} Z_{j,r}$. Risk factors are treated as independent across phases and within the reported ranges \citep{bge_zeitliche_2022}. This construction yields a flexible, data‐anchored representation of schedule risk: it preserves BGE’s activity-level judgment (probability and impact intervals) while avoiding over-parametrization.

We ran $N=300{,}000$ independent iterations. In each iteration, we draw $(T_2,T_3)$ and compute $T$. The simulated $T$ pins down the calendar for downstream milestones (construction, commissioning, emplacement, closure), which in turn yields a path of annual costs $\{C_t\}_{t=1}^{T}$ once inflation and nuclear-specific cost escalation. This timeline is then used as input to run the cost minimization model described in Section \ref{sec:minimization}.
This stochastic scheme directly maps governance-driven timing risk---as documented by BGE’s risk management for Phases~II and~III (see Table \ref{tab:phases}) and aligned milestone logic of ESK---into distributions of required financial performance and, if necessary, taxpayer backstops. It thereby links procedural delays to fiscal stress in a transparent and reproducible manner suitable for policy appraisal \citep{bge_zeitliche_2022,esk_verlangerte_2023}.

\subsection{Scenarios}

As noted in the introduction, several decades of delays have been projected for the German HLW repository site selection process. These anticipated delays necessitate the preparation of multiple disposal pathways to guarantee the safe and secure interim storage of nuclear waste in Germany \citep{eckhardt_trittsicherheit_2024}. In the present study, the assessment is restricted to three primary pathways that, despite differing time horizons, all ultimately aim to construct and operate a deep geological repository for HLW. Deep geological disposal is therefore assumed as the end state. As a result, the current situation is that a decision on the concrete pathway is not yet possible; instead, scenarios are defined and evaluated that reflect the range of plausible timelines and associated cost implications. 

Three scenarios are defined to assess the adequacy of KENFO’s fund under varying site selection and disposal timelines. This section first describes general assumptions applicable to all three scenarios before introducing the three scenarios themselves. Scenario~1 represents the legally prescribed timeline ("de jure", see Section \ref{sec:dejure}), serving as a reference case. Scenario~2 is a scenario bundle that introduces uncertainties by incorporating delays in the site selection process. The bundle of scenarios summarized as Scenario~3 extends the second scenario by considering a consolidated interim storage facility as an alternative approach. Each scenario's financial implications are analyzed using the previously described method (see Section \ref{sec:model}).

\subsubsection{General Assumptions}

Cost projections and data are primarily derived from \citet{wkgt_gutachtliche_2015}, in which costs are categorized into five categories: (i) decommissioning and dismantling, (ii) interim storage, (iii) costs associated with containers, transportation, and operational waste, (iv) the Konrad repository for LILW, and (v) the final HLW repository. Since the publication of the report in 2015, the organizational structure of Germany’s funding system has undergone significant changes. Specifically, financial responsibility for decommissioning and dismantling (i) has been transferred either to the former operators or, in the case of former GDR and research reactors, the state budget, see Section \ref{subsec:KENFO}. Consequently, these costs are excluded from this assessments of KENFO’s costs.

For interim storage costs (ii), estimates were sourced from utility balance sheets and internal financial data. Within the various scenarios considered in this study, these estimates were standardized and extended beyond the 40-year operational licensing period to ensure comprehensive cost projections for final waste disposal by 2088.

Transport and container costs (iii) include unit costs for storage containers, such as CASTOR and POLLUX for HLW, as well as MOSAIK containers for LILW. Additionally, this category accounts for transportation costs and expenses related to conditioning facilities required for packaging waste into these containers. The standardized cost data, provided by cask manufacturer GNS, were initially applied by individual utilities and later refined within the study to derive consistent cost estimates across different operational contexts.

Cost projections for the Konrad site (iv) were based on financial calculations from both the former Federal Agency for Radiation Protection BfS (now: BASE) and utilities.

Finally, cost estimates for the HLW repository (v) were derived from historical projections for the Gorleben site, adjusted for inflation and the utilities' cost-sharing responsibilities. Standardized recalculations were implemented to ensure consistency in financial projections.

Due to the absence of alternative sources, the cost projections in this work rely on \citet{wkgt_gutachtliche_2015}, which follows an engineering buildup methodology. This approach decomposes the national waste management program into a work breakdown structure (WBS), estimates individual component costs, and aggregates them into an overall financial projection for each fiscal year; the WBS is refined over time to improve accuracy.

Detailed cost projections were adjusted using actual expenditures reported in the annual federal budget, supplemented by financial data from official stakeholders such as BMUKN, BGE, and BGZ. Additionally, the timeline for each scenario was modified by adapting project structures in accordance with \citet{esk_verlangerte_2023} and \citet{bge_zeitliche_2022}. If the final disposal site selection is postponed by $n$ years, contingent liabilities for interim storage (e.g., maintenance, site security) increase, and replacement interim storage containers may be required. Overhead costs for CASTOR containers recur periodically, approximately once every 40~years, with each additional container incurring an estimated cost of €$_{2024}$2~m.; maintenance and site security costs are accounted for annually. 

A fixed inflation rate of 1.60\% and a nuclear-specific cost increase (NSCI) of 1.97\% are assumed for $n$ years under all scenarios (ceteris paribus) following the assumptions made by \citet{wkgt_gutachtliche_2015}. While our results are derived from a fund-adequacy perspective (ROI-driven compounding), the choice of an appropriate social discount rate for intergenerational nuclear-waste burdens is ethically contested; alternative social discounting assumptions (including low or zero rates as discussed by \citet{schulze_social_1981}) could change the interpretation of long-horizon cost burdens and should be treated as a key limitation and topic for future work. By holding macroeconomic factors constant, differences in projected outcomes are attributable to project-specific uncertainties (e.g., delays, operational risks). Another critical consideration is the time required for repackaging and transporting waste from interim storage facilities to the hypothetical final disposal site and the duration allocated for final disposal operations. These timeframes are each assumed to be at least 30 years, independent of the calculated delays (see \ref{sec:montecarlo}). Consequently, cost projections for the final disposal site have been extended to accommodate the revised timelines.

\subsubsection{Scenario 1: De Jure} \label{sec:dejure}
Scenario 1 serves as a reference case, outlining the timeline and key milestones for waste management activities as prescribed by existing legal and regulatory frameworks. No major deviations from the established schedule are assumed, and a baseline for comparison with alternative scenarios is provided \citep{esk_verlangerte_2023}. Figure \ref{fig:standAG_SOLL} in the annex shows this planned timeline.

In 2026, the extended interim storage approval process is initiated by BGZ, with regulatory inspection and approval by BASE \citep{base_overview_2019}. By 2027, provisions in §§~6 and~7 of the Atomic Energy Act apply to facilities such as Ahaus; storage beyond 40~years from the start of container storage is permissible only under exceptional circumstances, subject to parliamentary consultation \citep{bmj_gesetz_2022}. All BGZ facilities utilize dry storage for spent nuclear fuel (SNF), with loading scheduled for completion by 2028 \citep{esk_verlangerte_2023}. Beginning in 2032, the first containers reach technical expiration at several facilities (e.g., Ahaus, Lubmin, Krümmel, Biblis, Neckarwestheim); extended storage licenses must be granted by 2034. SNF remains at NPP sites until a permanent facility is operational; transport is expected to commence in 2045, continue for 30~years, and conclude interim storage activities by 2075 \citep{esk_verlangerte_2023}.

The process conducted by BGE proceeds in three phases, culminating in site selection (planned 2031), followed by approval processes for repository, treatment, storage systems, and final disposal containers (submissions by 2035; permits by 2040), enabling commissioning by 2045. From 2046, production of final repository containers and coquilles is carried out for 30~years; conditioning begins in 2048 (annual treatment of 340~Mg SNF and 130~glass coquilles). Disposal of waste in the repository itself is projected for 2050--2080 \citep{esk_verlangerte_2023}. This means that all to-be-funded activities end in 2080 as shown in \ref{fig:standAG_SOLL}. KENFO will not fund monitoring or maintenance costs after the repository is closed.

\subsubsection{Scenario 2: De facto}

The bundle of scenarios comprising "Scenario 2" is used to examine the financial impact of delays in the site selection process, which lead to extended interim storage and additional costs. To reflect these uncertainties, the timeline is adjusted using a stochastic model and Monte Carlo simulation as described in Section \ref{sec:model}. Extended or prolonged interim storage is assumed as a consequence of process delays communicated by BGE. The bundle incorporates modifications to the detailed breakdown of Scenario 1, with the adjusted timeline according to \citet{bge_zeitliche_2022} shown in Table \ref{tab:bge_risks}. Once the expected date for a final disposal site selection is established, all activities from Scenario 1 are systematically integrated into the revised timeline. These activities are assumed to occur as planned.

The simulation produces an earliest possible site selection in 2043, with a median of 2058, an average of 2052, and a latest possible conclusion in 2072. An additional 40~years must be added for the completion of all waste disposal activities. Relative to the ESK timeline, these results imply delays of 13 years at best and 36 years at worst. Such delays, result in recurring additional interim storage costs, primarily due to the necessity of replacing CASTOR and MOSAIK containers. For example, at Biblis and Neckarwestheim, the first round of container replacements is required in 2032, followed by a second cycle in 2072, and a third in 2112. The timing of replacements depends on project management strategies, particularly queue management. One potential optimization approach prioritizes transporting SNF from facilities with containers nearing end-of-life to the final disposal site before transporting from other facilities.

Beyond periodic overhead costs, continuous cost increases arise from prolonged maintenance and security measures. Additionally, cost projections for final disposal incorporate overheads that vary with the approach chosen in Phase~III---boreholes or exploratory mines. Borehole-based exploration (five years across six to ten regions) is significantly less costly and time-consuming than mine-based exploration (13 to 23~years across six to twelve regions), see Tables \ref{tab:phases} and \ref{sec:montecarlo}. These variations are incorporated through the adjusted schedules and cost categories defined above.

\subsubsection{Scenario 3: Consolidated Interim Storage}

Scenario~3 evaluates the feasibility of a centralized (or consolidated) interim storage facility (CISF) as an alternative to decentralized storage and assesses the financial and logistical implications of implementing such a facility before final disposal. It comprises a bundle of scenarios. The scenario assumes a single centralized facility that consolidates waste from multiple locations. It is assumed that, due to risk occurrences, delays in the site selection procedure, a CISF will be discussed in the 2050s, with a pathway switch to a CISF by 2080 \citep{eckhardt_trittsicherheit_2024}. A CISF may provide enhanced safety standards compared to multiple decentralized storage sites and could offer cost advantages due to economies of scale \citep{wegel_transporting_2019}. This scenario extends Scenario 2 by introducing two additional cost categories---(v) consolidated interim storage and (vi) transportation of wastes to consolidated interim storage---while all other variables remain unchanged.

The average cost (AC) per kilogram of heavy metal (kgHM) is modeled as a function of the facility’s total storage capacity in tons \citep{rothwell_spent_2021}. Offsite pool storage is considered instead of dry cask storage, since calculated maintenance and security costs are lower than container replacement costs given the prolonged duration of NWM activities relative to Scenarios~1 and~2. The cost estimation for offsite consolidated pool storage (converted using $\$_{2024}$1 $\approx$ €$_{2024}$0.93) and assuming capacities of 10{,}500~metric tons of heavy metal (MTHM) is

\begin{equation} \label{eq_CISF}
\begin{split}
AC_{\mathrm{off,pool}}=\left(176~\mathrm{\$_{2024}}+\frac{370{,}000~\mathrm{\$_{2024}}}{\mathrm{Capacity}_{\mathrm{off,pool}}}\right)\ \text{per kgHM}, \\
\Leftrightarrow AC_{\mathrm{off,pool}}=\left(164~\mathrm{\texteuro}_{2024}+\frac{344{,}100~\mathrm{\texteuro}_{2024}}{10{,}500}\right)=196.78~\mathrm{\texteuro_{2024}\ per\ kgHM}.
\end{split}
\end{equation}

This results in total estimated costs of approximately €$_{2024}$2.066~bn. These costs are then distributed across the specified timelines and appropriately discounted. As a consequence of these investments and timeline adaptation, final storage activities will be completed at later uncertain dates, thereby extending the closure of the repository well into the 22nd century. In this scenario, projected closure dates range from 2120 to 2180.

\section{Results} \label{sec:results}

An overview of the results is given in Table \ref{tab:results}. Note that all results are given in real values.

\begin{table}[h!]
\centering
\scriptsize
\begin{tabular}{p{3cm}p{3cm}p{3cm}p{3cm}} \hline
Scenario    & Total Discounted Cost {[}€\textsubscript{2024} bn.{]} & Minimum ROI   & Minimum Cash Injection at end of year 2024 (for fixed ROI at 3.7\%) {[}€\textsubscript{2024} bn.{]}  \\ \hline
1: De Jure  & 71                                    & 5.56\%        & 11.25                                                                                \\
2: De Facto & 111.22 - 277.02                                 & 5.91 - 6.35\% & 17.79 - 29.01                                                                        \\
3: CISF     & 520.46 - 783.13                                 & 5.54 - 6.63\% & 24.61 - 31.07 \\ \hline  
\end{tabular}
\caption{Overview of modeling results for all three scenarios.}
\label{tab:results}
\end{table}

\subsection{De Jure}
In the de jure scenario, the total discounted cost projection amounts to €$_{2024}$71~bn. This pathway follows the legally prescribed schedule, with site selection in 2031 and final disposal activities concluded by 2080.  

The results show that KENFO’s current fund volume is not sufficient to cover these projected costs when applying the statutory target return of 3.70\%. To achieve full cost coverage under the defined timeline, a minimum average ROI of 5.56\% is required. This value is substantially higher than the communicated target and therefore indicates that the fund is underfinanced if current assumptions remain unchanged.

If the ROI is fixed at 3.70\%, the existing fund balance would need to be increased by an additional lump sum capital injection of €$_{2024}$11.25~bn. to ensure adequacy through 2080. This adjustment represents the minimal upfront contribution necessary to compensate for the expected shortfall under the legal framework.  

The findings for the de jure case therefore demonstrate that the fund cannot achieve adequacy at the current target return. Either sustained higher investment returns or immediate taxpayer contributions are necessary to maintain solvency.

\subsection{De facto}
\begin{figure}[htbp]
  \centering
  \includegraphics[width=0.85\textwidth]{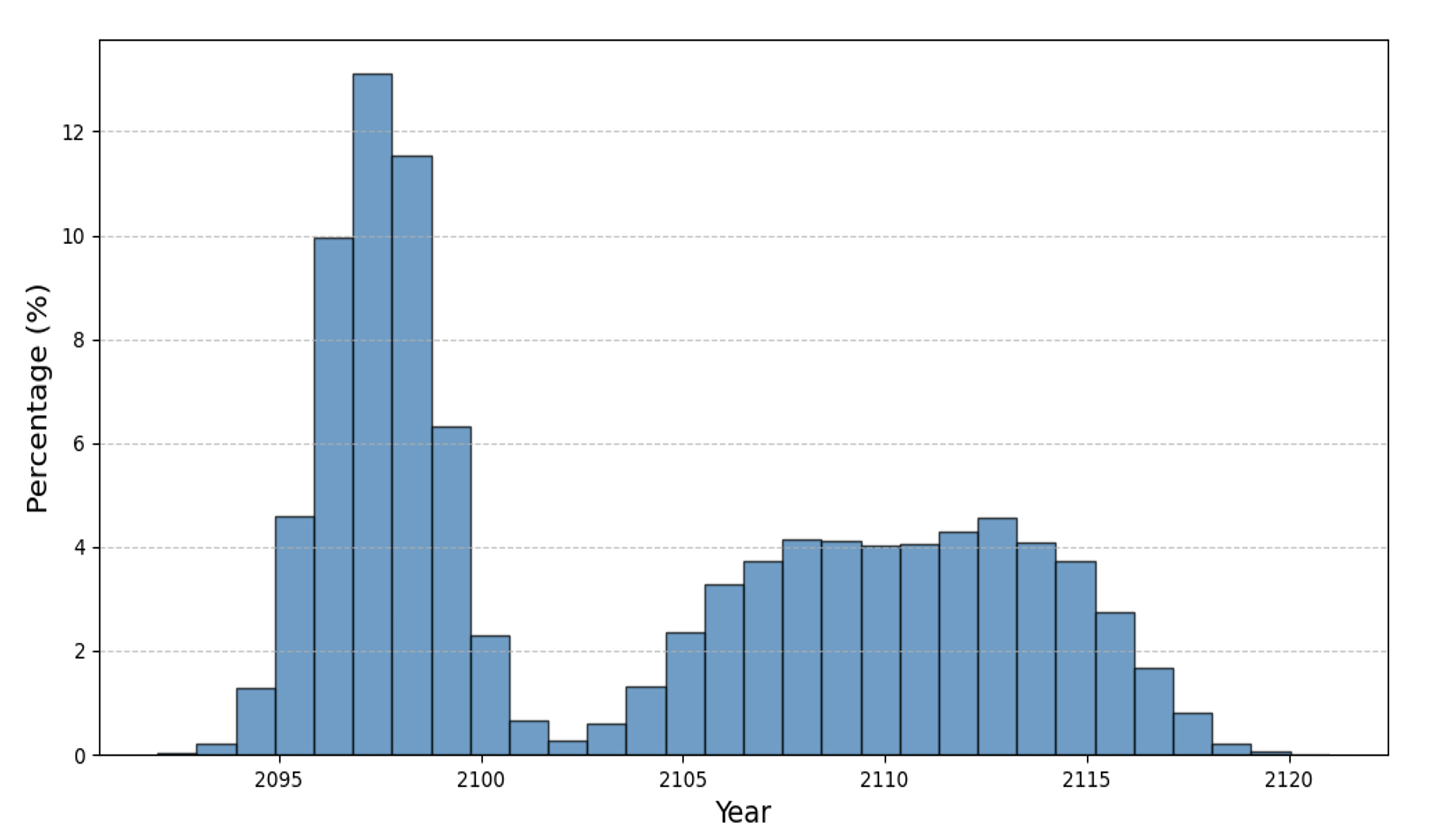}
  \caption{Histogram of closure dates of a final repository in Scenario~2 for all 300,000 simulation runs.}
  \label{fig:hist_fdsaconclusion}
\end{figure}

Scenario 2 represents a probabilistic framework that accounts for multiple potential timelines, spanning a projected duration of 28 years for the completion of the site selection procedure. Our stochastic simulation produces a distribution of possible completion years for the final disposal site activities, spanning from 2093 to 2120. Figure~\ref{fig:hist_fdsaconclusion} illustrates this distribution, revealing a distinct left-skewed bimodal pattern with a primary peak between 2092 and 2102 and a smaller secondary peak between 2102 and 2120. This distribution results mostly from the model's selected underground exploration technique in Phase~III of the site selection procedure. Exploration via deep boreholes is estimated to require 5 to 6 years, while exploratory mines could require 13 to 23 years.

\FloatBarrier

\begin{figure}[htbp]
  \centering
  \includegraphics[width=0.85\textwidth]{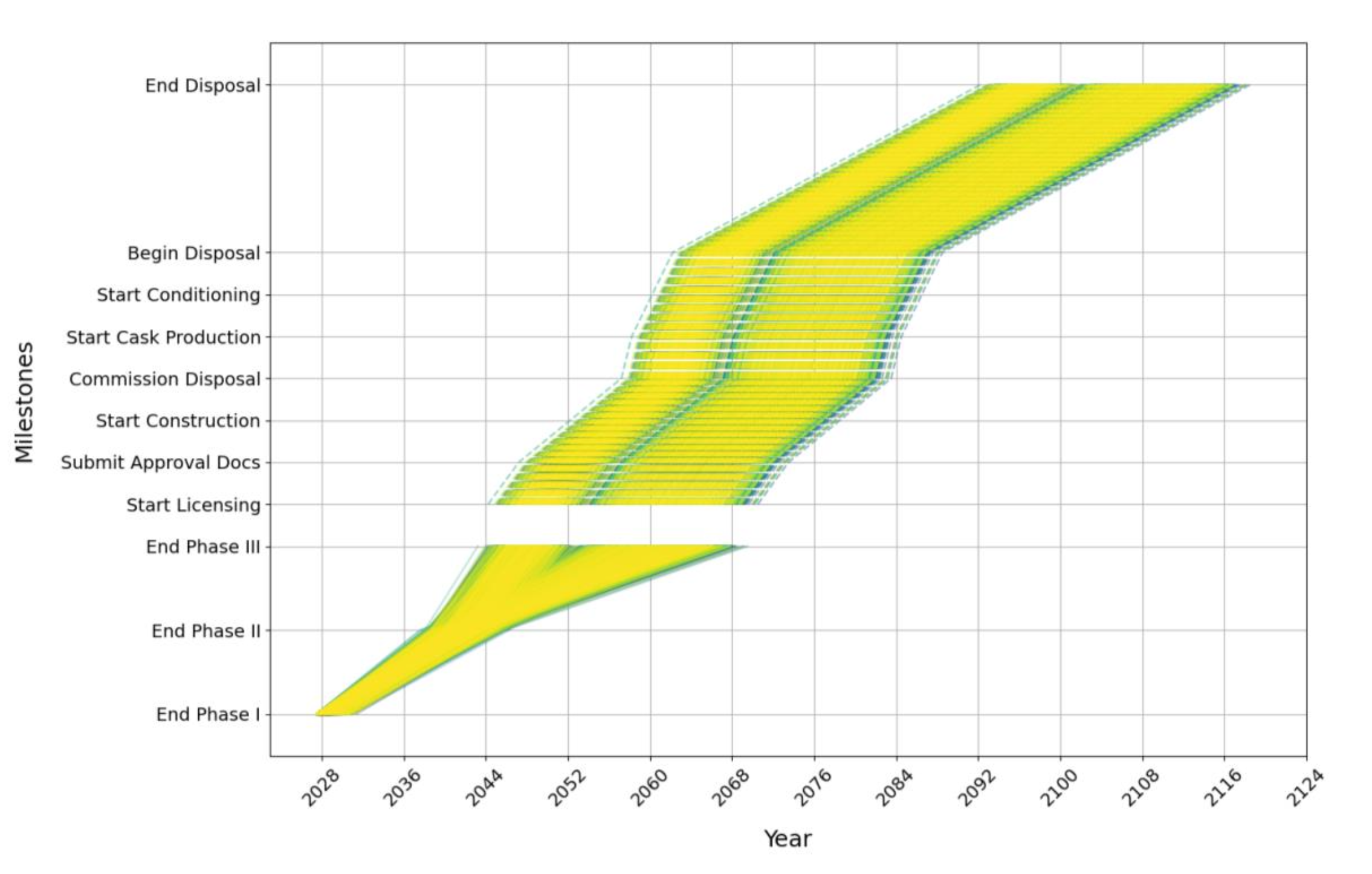}
  \caption{Range of nuclear waste management timeline and corresponding milestones of Scenario~2.}
  \label{fig:timeline_de_facto}
\end{figure}

The main milestones and phases to be carried out by  BGE in Scenario~2 are shown in Figure~\ref{fig:timeline_de_facto}, which lists milestones on the y-axis and the projected timeframe for their completion on the x-axis. For visual simplicity, the graph begins in 2028 (after the assumed completion of Phase~I) and ends in 2120, as the last conclusion of a simulated reality takes place in 2120.

\FloatBarrier

Under no sub-scenario of Scenario~2 does KENFO’s fund volume (as of 2024) suffice to meet projected costs at the statutory target ROI of 3.70\%. Instead, the minimum average ROI required for full cost coverage increases to 5.91\% for the lower bound of the timeline (2093) and rises to 6.35\% for the upper bound (2120).

\begin{figure}[htbp]
  \centering
  \includegraphics[width=0.85\textwidth]{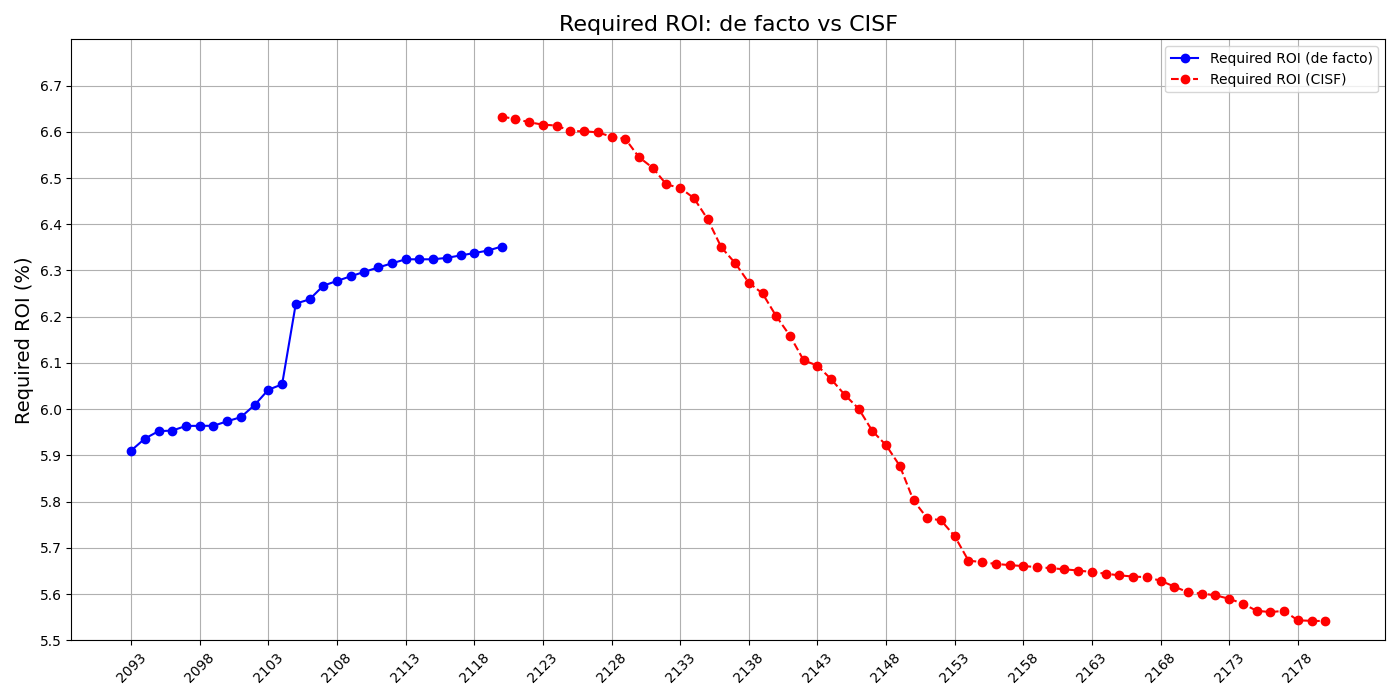}
  \caption{Required annual average ROI of Scenario~2 and 3.}
  \label{fig:roi_cisf}
\end{figure}

Figure~\ref{fig:roi_cisf} shows the required long-term average ROI across the projected timeframes for Scenarios~2 and~3. The y-axis indicates the year of repository closure. A steeper increase in the required ROI is observed between 2104 and 2105, attributable to periodic replacement costs for CASTOR and MOSAIK containers.

\FloatBarrier

\begin{figure}[htbp]
  \centering
  \includegraphics[width=0.85\textwidth]{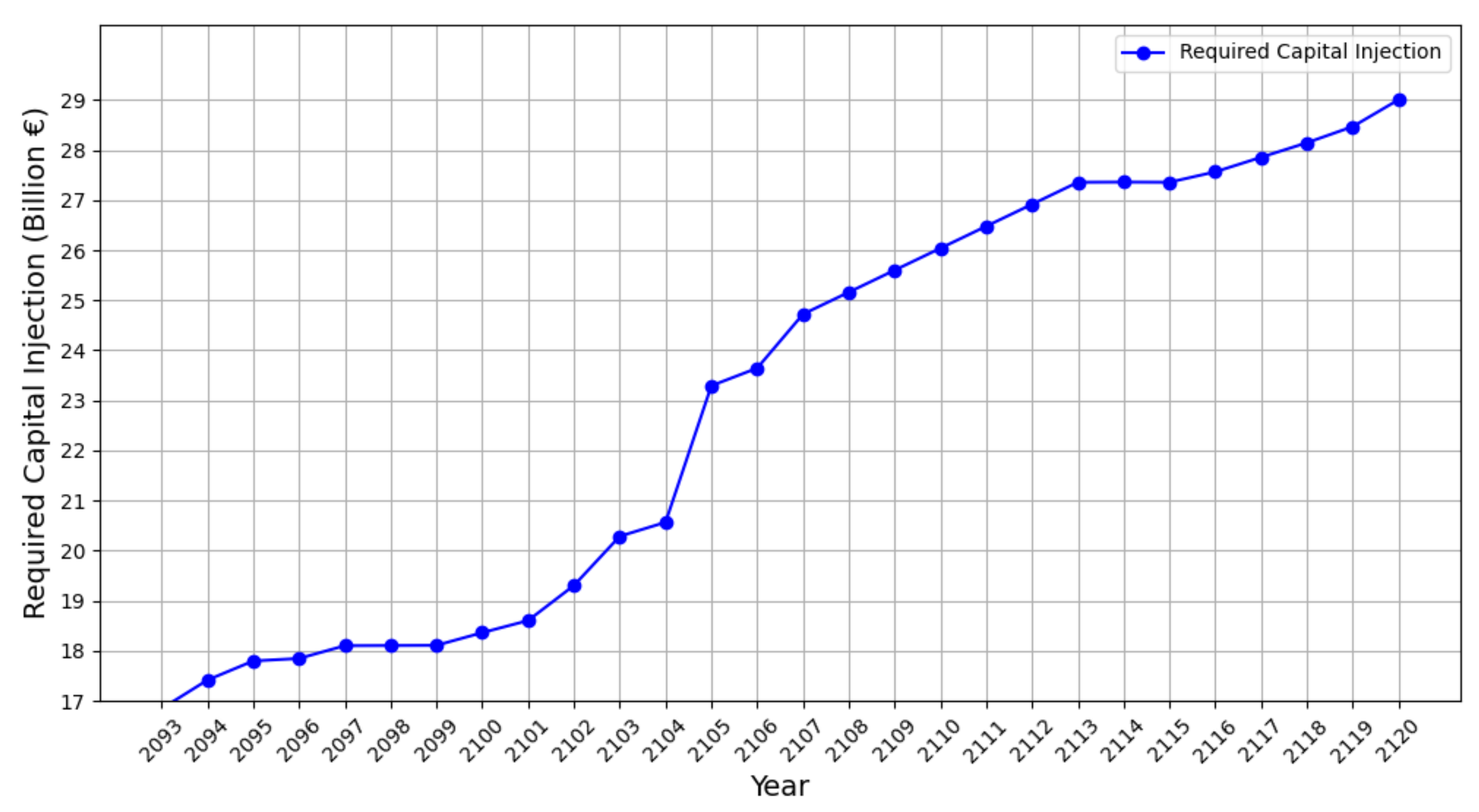}
  \caption{Required lump sum capital injection of Scenario~2 for different repository closure years. \newline Note: The required lump sum is to be deposited as a one-time payment at the end of 2024.}
  \label{fig:injection_de_facto}
\end{figure}

If the ROI is fixed at 3.70\%, lump sum capital injections are necessary to compensate the shortfall. For the earliest and latest sub-scenarios (repository closed in 2093 and 2120), €$_{2024}$17.79~bn., and €$_{2024}$29.01~bn, are required, respectively. Figure~\ref{fig:injection_de_facto} illustrates the required lump sum capital injections across all sub-scenarios.

\FloatBarrier

These results demonstrate significant financial stress in Scenario~2 due to extended interim storage, recurring container replacements, and prolonged timelines. The current target ROI is not sufficient to fund the process in its current design.

\subsection{Consolidated Interim Storage}

The bundle comprising Scenario~3 indicates that the minimum average ROI required to cover projected costs is 6.63\% for the sub-scenario with repository closure in year 2120. This percentage gradually declines in subsequent years, reaching the upper bound of the scenario consideration in 2180 with an average return on investment of 5.54\%, as shown in Figure~\ref{fig:roi_cisf}.  

If KENFO’s target return of 3.70\% is maintained, the fund’s current value would have to be increased to €$_{2024}$31.07~bn. in the first sub-scenario and €$_{2024}$24.61~bn. in the last sub-scenario by the end of 2024. Notably, an inverse relationship exists between the required return on investment or capital infusion and the projected timeframe, with later years requiring lower returns or smaller lump sum capital injections for each sub-scenario. This results from the reduced operational costs and reduced container refurbishments if all waste is transferred to a CISF.

\section{Discussion} \label{sec:discussion}

Long-term models are built on a range of assumptions and uncertain inputs and are thus inherently subject to uncertainty. To fully understand their structure and behavior, it is essential to examine how these uncertain factors influence the model's outcomes \citep{saltelli_sensitivity_2002}. In the following section, the results of a sensitivity analysis are presented. Moreover, it discusses the limitations inherent to the proposed model.

\subsection{Sensitivities}

This section explores the influence of varying one independent variable (inflation rate) on the required ROI and lump sum capital injection levels.

In this model, there are four independent variables---ROI, NSCI, project timelines, and the decision to implement a CISF---that affect the minimum required ROI and the lump sum capital injection in 2024. We limit our sensitivity analysis to the inflation rate that we assume as 1.6\% based on \citet{wkgt_gutachtliche_2015}--an optimistic value given inflation rates in recent years. We thus adjust the inflation rate to 1.72\%, 2.02\%, and 3.70\%. These rates reflect distinct economic conditions observed during different recent historical periods, i.e., the average inflation rate in Germany over 1991--2021, the average inflation rate from 1991 to 2024, including the COVID19-pandemic and Russia's war in Ukraine, and the average inflation rate over the last five years. These scenarios provide a comprehensive view of how different inflation assumptions influence the calculated minimum required ROI, with the analysis focusing exclusively on the resulting effect on the minimum ROI.

In Scenario~1, the minimum required ROI increases as the inflation rate rises, with approximately 5.60\% at an inflation rate of 1.60\%, 6.18\% at 1.72\%, 7.56\% at 2.02\%, and 10.54\% at 3.70\%, respectively. This trend indicates that higher inflation rates necessitate progressively greater annual returns to cover projected costs and maintain long-term financial adequacy.

\begin{figure}[htbp]
  \centering
  \includegraphics[width=0.85\textwidth]{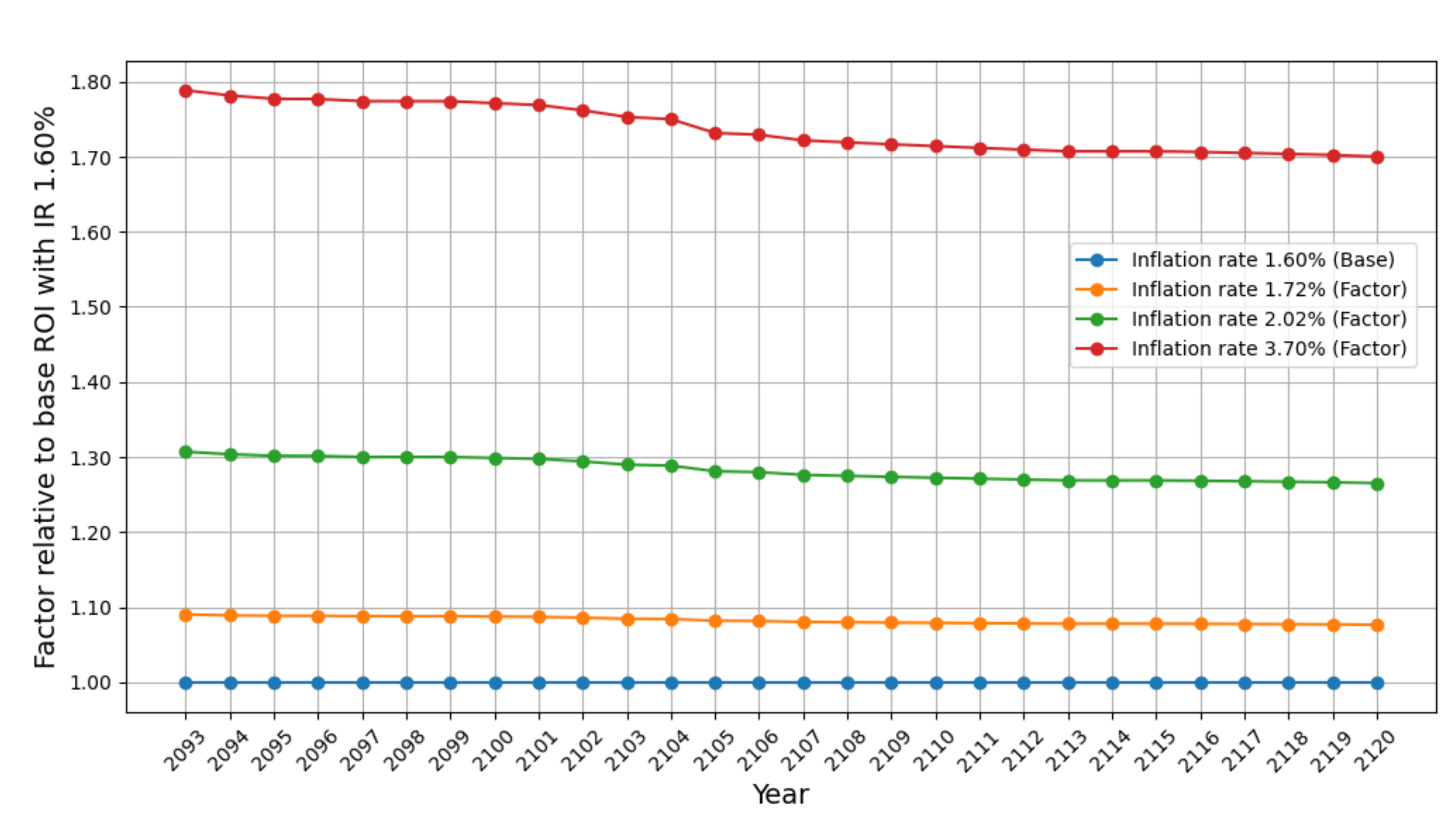}
  \caption{Sensitivity analysis results of Scenario~2 bundle. \newline Abbreviations: ROI = Return on Investment; IR = inflation rate. Source: Own depiction.}
  \label{fig:sens_s2}
\end{figure}

As shown in Figure~\ref{fig:sens_s2}, higher inflation rates require higher ROIs but the relation is non-linear: over time, higher inflation rates are negated by the compound interest effect. The base inflation rate of 1.60\% results in ROIs ranging from approximately 5.91\% in 2093 to 6.35\% in 2120 (average \(\sim 6.13\%\)). For 1.72\% inflation, the ROI ranges from roughly 6.44\% to 6.84\% (average \(\sim 6.64\%\)), i.e., a average factor of \(\sim 1.10\). For a 2.02\% inflation rate, the ROI ranges from 7.72\% to 8.04\% (average \(\sim 7.88\%\)), a factor of \(\sim 1.30\) relative to the base case. At 3.70\% inflation, it varies from approximately 10.57\% to 10.80\% (average \(\sim 10.68\%\)), which corresponds to a factor of \(\sim 1.75\).

This sensitivity analysis shows the risks of assuming overly optimistic inflation rates for long-term financial analyses.

We conduct an analogous analysis for Scenario~3. For the base case, the ROI decreases from a maximum of approximately 6.63\% in 2120 to a minimum of about 5.54\% in 2180 (average \(\sim 6.09\%\)). For an 1.72\% inflation rate, the absolute required ROI ranges from about 5.91\% to 7.15\% (average \(\sim 6.53\%\)), which equals a factor of \(\sim 1.08\) relative to the base. At 2.02\%, the ROI spans roughly 6.80\% to 8.43\% (average \(\sim 7.62\%\)), i.e., a factor of \(\sim 1.25\). At 3.70\%, it varies from around 8.82\% to 11.32\% (average \(\sim 10.07\%\)), implying a factor between \(\sim 1.50\) and \(\sim 1.75\). Over time, we find that the compound interest on ROI compensates the negative influence of higher inflation rates.

Overall, the sensitivity analysis shows that, depending on realized inflation over the next years, KENFO might need to achieve a higher ROI by a factor of up to \(\sim 1.8\). Such a high ROI (\(\sim 6.0\%\)) does not align with common low-risk long-term investment strategies (e.g., AAA-rated bonds at roughly 1--2\%~p.a.).

\subsection{Limitations}

This analysis considers the financial implications for KENFO's ability to fund the German radioactive waste disposal process. The stochastic model, however, is limited to the potential delays in the site selection procedure for a final repository for HLW and neglects other potential delays, such as delays in the construction of the repository itself, technical challenges in providing suitable containers and the transport of waste, the influence of protests, delays at the Konrad mine, whose costs are also covered by KENFO, and non-foreseeable events. Our assumptions are based on risk assessments by \citet{bge_zeitliche_2022} that are considered as overly optimistic by \citet{krohn_unterstutzung_2024}.

Given the lack of reliable cost data for waste disposal activities, our cost assumptions are based on a study from 2015 that in turn is based on an outdated repository design \citep{wkgt_gutachtliche_2015}. We acknowledge that these cost assumptions are likely too low.

Furthermore, our analysis assumes conservative NSCI values of 1.97\% and a low inflation rate of 1.6\% that lead to lower ROI and cash injection requirements. Discounting practices can further skew the impression of such a long-term analysis. \citet{schulze_social_1981} advocate for discount rates of 0\% to acknowledge the interests of future generations. Such an analysis and its implications on nuclear waste financing should be subject to future work.

This study is also explicitly not concerned with KENFO's investment strategy. Currently, KENFO holds a significant part of its wealth in easily liquidizable assets and is waiting for opportunities to invest in long-term, more secure, and potentially revenue-generating, non-liquid, assets such as housing \citep{kenfo_geschaftsbericht_2025}. A consideration of KENFO's investment strategy and an analysis of its compliance with ESG criteria could be subject to future work.

\section{Conclusion and Policy Recommendations} \label{sec:conclusion}

With the closure of its last commercial nuclear power plants Germany is one of the few countries whose radioactive waste inventory will not increase over time--the identification and construction of a waste repository for HLW should therefore be easier than in other countries who will have to take into account additional wastes from extended operations or new power plants. Regardless, the German process is already projected to be delayed by several decades after having been redesigned less than a decade ago with an ambitious target of identifying a suitable location for an HLW repository by 2031.

Therefore, this analysis considers the implications for the sufficiency of nuclear waste fund KENFO to ensure funding for the disposal of Germany's radioactive wastes under consideration of currently projected delays that will place the closure of a future repository far into the 22nd century.

Depending on the assumed scenarios, total cost projections range from €$_{2024}$71 to 804~bn., requiring average ROIs over the next decades ranging from 5.56\% and 6.63\%---well above KENFO’s target of 3.70\%. These results assume optimistic cost data, no further delays in the site selection procede---apart from those acknowledged by BGE---and low inflation rates. Assuming KENFO achieves its average target of 3.70\% ROI, the fund would have required capital injections of €$_{2024}$11.25~bn.  to €$_{2024}$31.07~bn. at the end of 2024.

The burden on future generations, that will have to organize the physical process of waste disposal, should be reduced at least in terms of the financial aspects. In general, German policy-makers have two general options to ensure KENFO's fund adequacy.

First, one could attempt at reducing the costs of the process. This could be achieved by accelerating the site selection process by fostering experience from other countries, such as France, and Finland, increased international collaboration in cask and repository design studies, or by following a more agile approach of parallel surface and underground explorations and focusing on one type of host rock.

Second, KENFO could attempt at increasing its revenue. This can either be achieved through higher ROIs which are usually earned via higher-risk and potentially ESG-non-compliant investments, or by increasing KENFO's volume today, as proposed by our analysis. Increasing KENFO's volume from today's state budget could shift some of the additional expected financial burden from future generations towards current taxpayers who profited from the operation of nuclear power plants in Germany.

Overall, we find that the German waste disposal process and its funding remain uncertain. It is highly likely that KENFO will not be able to provide sufficient funds for the whole process. Consequently, it is essential to pro-actively discuss potential fall-back options and final liabilities, such as waste levies on electricity prices or funding via the state budget \citep{wimmers_finanzierungsfragen_2025}, which should be subject to future research.

To conclude, a proactive discussion of the financial implications of KENFO's insufficiency could be a first step towards the acknowledgment of this additional burden that is being placed on future generations through the current delays of the site selection procedure. 


\section*{Data Availability}
The model source code and all relevant data is accessible at \url{https://github.com/Mahdi-Awawda/KENFO/tree/main}.

\section*{Author Credit Statements}
M.A.: Methodology, Formal Analysis, Data Curation, Investigation, Software, Visualization, Writing - Original Draft; A.W.: Conceptualization, Methodology, Investigation, Validation, Writing - Original Draft.

\section*{Declaration of Interests}

The authors have no known competing interests to declare.

\section*{Acknowledgements}

We thank the attendees of the 2025 AT-OM Day at TU Berlin, and those of the 3rd safeND Symposium 2025 for their constructive input on this work. We extend our gratitude to Fanny Böse, Friederike Frieß and Christian von Hirschhausen for their helpful comments on earlier drafts of our manuscript. This work is a methodological advancement of earlier work \citep{wimmers_very_2026}.


\bibliographystyle{elsarticle-num-names}
\bibliography{anyatom_references.bib}

\appendix

\section{German Waste Disposal Process}

\begin{figure}[htbp]
  \centering
  \includegraphics[width=0.85\textwidth]{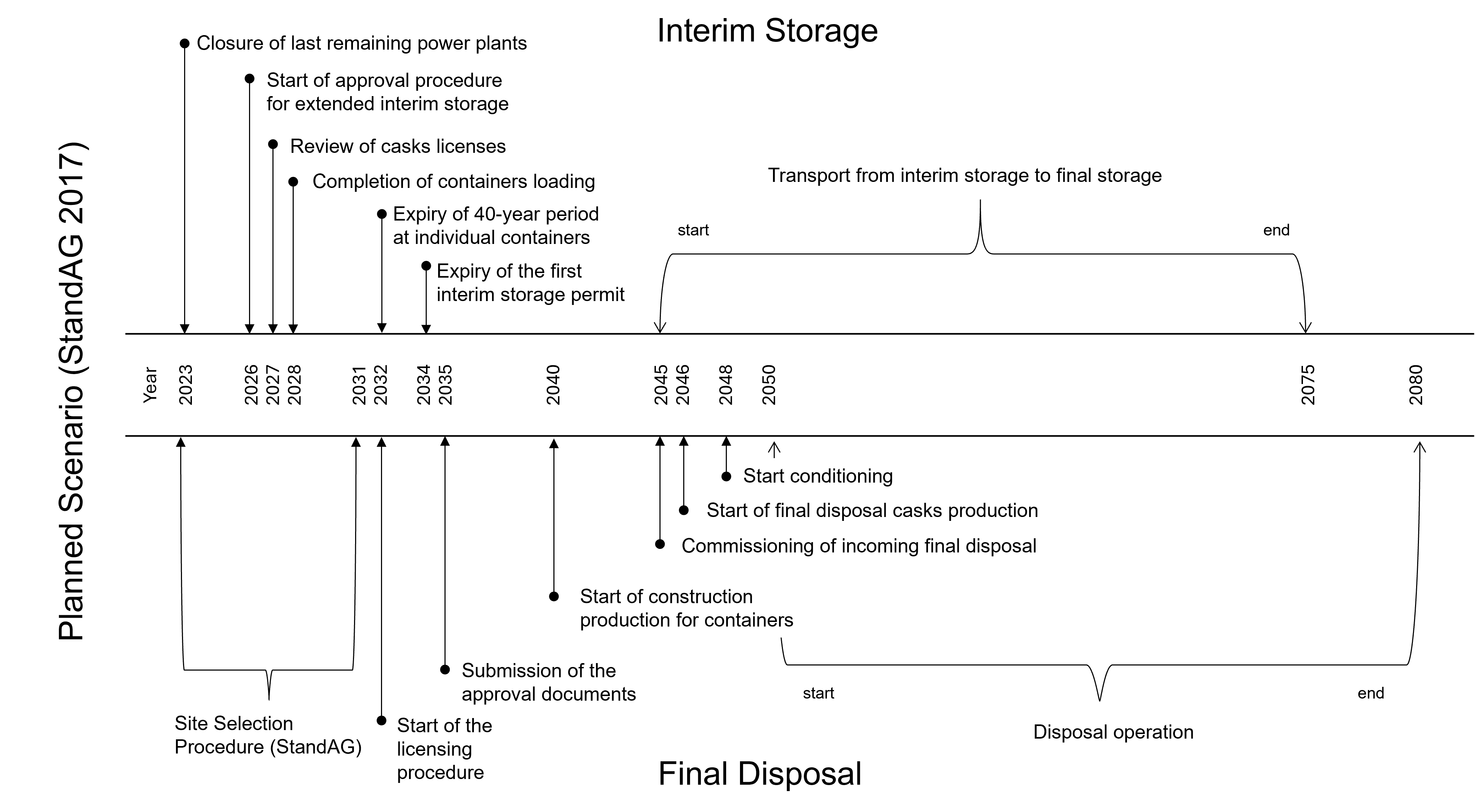}
  \caption{Duration of German nuclear waste disposal activities according to StandAG. Source: Own depiction with data taken from \citet{esk_verlangerte_2023}.}
  \label{fig:standAG_SOLL}
\end{figure}

\section{Input for Monte Carlo Simulation}

\begin{center}
\scriptsize
\begin{longtable}{p{7cm}p{1cm}p{1cm}p{1cm}p{1cm}}
\caption{Events with corresponding probabilities and delays following \citet{bge_zeitliche_2022}.} \label{tab:bge_risks} \\

\hline \multicolumn{1}{c}{\multirow{2}{*}{\textbf{Risks   and Measures}}} & \multicolumn{2}{c}{\textbf{Probability of Occurrence}} & \multicolumn{2}{c}{\textbf{Estimated Time Impact (Months)}}  \\
\multicolumn{1}{c}{}  & \textbf{Min} & \textbf{Max}  & \textbf{Min} & \textbf{Max}  \\
\hline 
\endfirsthead

\multicolumn{5}{c}%
{{\tablename\ \thetable{} -- continued from previous page}} \\
\hline \multicolumn{1}{c}{\multirow{2}{*}{\textbf{Risks   and Measures}}} & \multicolumn{2}{c}{\textbf{Probability of Occurrence}} & \multicolumn{2}{c}{\textbf{Estimated Time Impact (Months)}} \\
\multicolumn{1}{c}{}  & \textbf{Min} & \textbf{Max}  & \textbf{Min} & \textbf{Max}  \\ \hline 
\endhead

\hline \multicolumn{5}{r}{{Continued on next page}} \\ \hline
\endfoot

\hline \hline
\endlastfoot

Procedure   (method) for the representative preliminary safety   investigations is not compliant with legal requirements                                                                                                                                                                                & 5\%                        & 5\%                       & 6                            & 12                           \\
Serious   dissent following public discussion of the method for carrying out the representative preliminary safety investigations                                                                                                                                                                       & 5\%                        & 5\%                       & 2                            & 6                            \\
Data   deliveries are delayed and/or not in the desired quality                                                                                                                                                                                                                                         & 50\%                       & 50\%                      & 2                            & 6                            \\
Favorable   areas within the BGE category A areas are overlooked                                                                                                                                                                                                                                        & 20\%                       & 20\%                      & 12                           & 24                           \\
Need   to further develop methodology for renewed application of Geoscientific   Assessment Criteria (geoWK) (§24 StandAG) (methodological/technical)                                                                                                                                                   & 5\%                        & 5\%                       & 6                            & 12                           \\
Methodological   difficulties with the application of Geoscientific   Assessment Criteria (geoWK) (§24 StandAG) are underestimated                                                                                                                                                                      & 5\%                        & 5\%                       & 6                            & 12                           \\
Serious   dissent with the public on the proposed method for determining siting regions   according to §13 StandAG in Phase 1 Step 2 of the site selection procedure                                                                                                                                    & 5\%                        & 5\%                       & 6                            & 12                           \\
Insufficient   plausibility of figures on quantities from TUR as input and output to representative preliminary safety investigations, application   of Geoscientific Assessment Criteria (geoWK) and Planning-Scientific   Assessment Criteria (planWK) values that are considered massively excessive & 30\%                       & 30\%                      & 6                            & 12                           \\
Requirements   for further exploration differ from those assumed by the BGE                                                                                                                                                                                                                             & 30\%                       & 30\%                      & 2                            & 6                            \\
Consideration   of the results of the expert conference on partial areas ("Fachkonferenz   Teilgebiete") is not comprehensible with regard to the BGE’s proposal   for site regions                                                                                                                     & 5\%                        & 5\%                       & 6                            & 12                           \\
Possible   delays in the granting of permits according to the Federal   Mining Act (BBergG)                                                                                                                                                                                                             & 70\%                       & 80\%                      & 6                            & 12                           \\
Possible   delays due to conflict mediation with citizens, when exploration measures of   the BGE are affected                                                                                                                                                                                          & 50\%                       & 60\%                      & 6                            & 12                           \\
Bottlenecks   in exploration and drilling capacity: Site exploration under StandAG sets a   demand peak over a relatively short period. This can lead to bottlenecks in   exploration service providers.                                                                                                & 30\%                       & 40\%                      & 6                            & 12                           \\
For   individual works (e.g., hydraulic test work in the borehole) special   limitations exist                                                                                                                                                                                                          & 20\%                       & 30\%                      & 6                            & 12                           \\
The   size of the site regions will significantly impact the required timeframe of   exploration                                                                                                                                                                                                        & 20\%                       & 30\%                      & 6                            & 12                           \\
There   may be significant delays (0.5 to 1 year per site region) when processing 3D   seismic data                                                                                                                                                                                                     & 60\%                       & 80\%                      & 6                            & 18                           \\
Missing   or expiring permits (§16 StandAG)                                                                                                                                                                                                                                                             & 20\%                       & 20\%                      & 6                            & 12                           \\
Obtaining   use and easement rights from landowners takes too long for surface   exploration                                                                                                                                                                                                            & 30\%                       & 50\%                      & 3                            & 6                            \\
Possible   delays in the granting of permits according to the Federal   Mining Act (BBergG)                                                                                                                                                                                                             & 30\%                       & 50\%                      & 6                            & 12                           \\
Time   delays due to conflict mediations with citizens, when exploration measures of   the BGE are affected                                                                                                                                                                                             & 50\%                       & 60\%                      & 6                            & 12                           \\
Bottlenecks   in exploration and technical capacities: Site exploration under StandAG   creates a demand peak that exceeds available capacities                                                                                                                                                         & 20\%                       & 30\%                      & 6                            & 12                           \\
Missing   or expiring permits (§18 StandAG)                                                                                                                                                                                                                                                             & 30\%                       & 50\%                      & 6                            & 12                           \\
Obtaining   use and easement rights from landowners (permitting) for surface exploration   takes too long                                                                                                                                                                                               & 30\%                       & 50\%                      & 3                            & 6                           

\end{longtable}
\end{center}

\end{document}